\documentstyle[12pt,aasms4]{article}
\lefthead{Whiting}
\righthead{Least-Action Theory}
\begin{document}

\noindent \today

\title{The Least-Action Principle: Theory of Cosmological Solutions
and the Radial-Velocity Action}
\author{Alan B. Whiting}
\affil{ Physics Department, U. S. Naval Academy}
\authoraddr{Annapolis, MD 21403}

\begin{abstract}
Formulating the equations of motion for cosmological bodies (such as
galaxies) in an integral, rather than differential, form has several
advantages.  Using an integral the mathematical instability at early
times is avoided and the boundary conditions of the integral correspond
closely with available data.
Here it is shown that such a least-action calculation
for a number of bodies interacting by gravity has a finite
number of solutions, possibly only one. 
Characteristics of the different
possible solutions are explored.  
The results are extended to cover the motion of a continuous fluid.
A method to generalize an action to use velocities, instead of
positions, in boundary conditions, is given, which reduces in
particular cases to those given by Giavalisco et al. (1993)
\markcite{G93} and  Schmoldt \& Saha (1998) \markcite{SS98}.
The velocity boundary condition is shown to have no effect on
the number of solutions.
\end{abstract}

\section{Introduction}

The present motions of cosmological objects, in particular galaxies,
are functions of their past history.  In principle one might
discover the shape of the past by calculating presently observed
positions and motions backward.  However, in doing this 
 we are faced immediately with two problems.  First,
their velocities in the plane of the sky are not known, and their distances
not known accurately; so perhaps half the information needed to start 
the calculation by Newton's equations is there.  Second, if the trajectories
of the galaxies are to be traced back to very early times distances become
very small and corresponding gravitational forces very large.  Small errors
in present velocities or positions become heavily magnified, resulting in
galaxies being formed at infinite speeds.  The problem is mathematically
unstable, rather like trying to roll a marble to the top of a glass
mountain, and requiring that it stop exactly on the summit\footnote{Valtonen 
et al. \markcite{V93} (1993) have found some possible solutions for
the motion of the major galaxies in the Local Group and the 
Maffei 1/IC 342 Group by integrating equations of motion forward from
an early time.  However, it is not clear that this method is generally
applicable, and in any case requires a great deal of hunting about in
parameter space; for their result, the Valtonen group integrated ten
thousand situations.}.

To avoid these difficulties Peebles (\markcite{P89}1989, \markcite{P90} 
1990, \markcite{P94} 1994) 
formulated the problem in integral
rather than differential form.  This traded the relative simplicity and
definiteness of differential equations for the stability of the integral.
The most important consideration in moving from the differential to 
integral form of the problem (apart from the mechanics of implementation)
is the fact that, with the same boundary conditions, an integral calculation
may produce several (or many) solutions.  An obvious question to answer
is just how many there are.  This is something more than a purely mathematical
concern.  Of course, if the numerical calculation of solutions can be
guided in some way there is the potential for a large savings in computer
time, and if the number of solutions is limited the search may be stopped
when all are found.  Conversely, if the number of solutions is very large
or infinite, the usefulness of the calculation is thrown into doubt
(unless some method of selecting more probable solutions is found).  But
the question is more fundamental than that, for the variational formulation of
the cosmological problem corresponds closely to the limits of our knowledge.
When the present radial velocities and positions on the sky
of a number of bodies are specified and the
Big Bang postulated, 
we find the end conditions are fixed; the action is determined by
relevant physics.  The mathematical question is thus transformed
into a cosmological one.

The subject of this study is the mathematical theory of variational
calculations as applied to the cosmological problem.  That problem is
defined as the determination of 
the motion of a number of bodies moving under gravitational interaction,
with the requirement
that all bodies must be at the same point (in proper coordinates)
at $t=0$. Newtonian, rather than relativistic, calculations
are employed throughout\footnote{See Peebles \markcite{P80} (1980) and
Bondi \markcite{B60} (1960) for the validity of this approach.}.

The cosmological problem may be interpreted as a rough approximation of the
motion of galaxies, each galaxy simulated by a point mass interacting
only through gravity.  This is the way most least-action calculations have
proceeded, and is not a bad approximation considering the uncertainties
in such data as distances and masses.  It would be more accurate, however,
to consider the objects to represent the dark matter halos of galaxies (which
as far as is known interact {\em only} through gravity).  The point-mass
approximation provides a reasonable simulation of gravitational effects,
since multipole moments decay rapidly with distance (Dunn \& Laflamme 1995
found them to be quite unimportant), and in any case the conclusions of
this study are not affected by the detailed form of the gravitational
potential used.

Of course, identifying whole galaxies with single bodies ignores internal
structure (which may be significant in some cases) and the effects of
mergers (which certainly are significant);  Dunn \& Laflamme (1995)
found some additional problems.  To address these matters one
must turn to a continuous fluid formulation of the problem.  Section 6
generalizes the discrete-body results to this more complicated situation.
 
\section{The First Variation}

Consider first the problem of minimizing the integral 
\begin{equation}
I = \int_{t_0}^{t_1} \left( T - V \right) dt
= \int_{t_0}^{t_1} L \left( q_i,\dot{q_i},t \right) dt
\label{eq:integral}
\end{equation}
where the kinetic energy $T$ is quadratic in the generalized velocities
$\dot{q_i}$ (or, alternatively, in the generalized momenta $\partial L /
\partial \dot{q_i} = p_i$) and the potential $V$ does not depend on velocity.
The end points $q_i(t_0)$ and $q_i(t_1)$  are given.  
(This is interpreted dynamically
by constructing a path ${\bf r}(t)$
in 3n-dimensional space
using the vectors ${\bf r}_j(q_i(t))$, $j =1$ to $n$, $i=1$ to $3n$,
where the ${\bf r}_j$ are
the paths of the $n$ bodies in 3-dimensional space.)
For small variations
the change in the action is given by a truncated expansion in a Taylor's
series (treating $q_i,\dot{q}_i$ as independent variables):
\begin{equation}
\delta \int_{t_0}^{t_1} L \left( q_i,\dot{q_i},t \right) dt =
\int_{t_0}^{t_1} \sum_i \left[ \frac{\partial L}{\partial  q_i}
\delta  q_i +
\frac{\partial L}{\partial \dot{q}_i}  \delta \dot{q}_i
\right] dt = 0. 
\label{eq:variation}
\end{equation}
Equation
(\ref{eq:variation}) can be integrated by parts to give
\begin{equation}
\int_{t_0}^{t_1} \sum_i \left[ \frac{\partial L}{\partial  q_i} -
\frac{d}{dt} \left( \frac {\partial L}{\partial \dot{q}_i} \right)
\right]  \delta  q_i dt  + \sum_i
\left[ \frac {\partial L}{\partial \dot{q}_i}  \delta  q_i
\right]_{t_0}^{t_1} = 0
\end{equation}
and if the variation in ${\bf r}$ vanishes at the end points (that is, if
the end points are fixed) the boundary term is zero.  The requirement
that the integral
vanish for arbitrary  variations $ \delta {\bf r}$ results in 
the Euler-Lagrange equations:
\begin{equation}
\frac{\partial L}{\partial  q_i}
-\frac{d}{dt} \left( \frac {\partial L}{\partial \dot{q}_i} \right)
=0.
\end{equation}
These are the dynamic equations, identical with those derived from (for
example) forces and accelerations.  The correspondence between the
dynamic equations and the vanishing of the first variation of equation 
(\ref{eq:integral}) is {\em Hamilton's Principle}.

A path which minimizes the integral will thus always satisfy the
dynamic equations.  However, the converse is not necessarily true:
a path satisfying the Euler-Lagrange equations is not guaranteed
to provide a minimum of the corresponding integral.  For sufficiently
small path lengths a minimum does result (see, for example, Whittaker 
\markcite{W59} 1959,
pp. 250-2); beyond a certain point the path 
will make the integral stationary, but not necessarily a minimum.
Finding the point that determines the limit of application
of the least-action technique (strictly interpreted)
to the dynamical problem will be discussed
below\footnote{Whittaker calls Euler's Principle, which appears later,
the {\em Least Action} Principle; however, Hamilton's Principle has
also been called this.  To distinguish between the two I will use
the names of the mathematicians and call them collectively Least Action
Principles.}.

Adding a total derivative (in multiple dimensions, a divergence expression)
will not change the Euler-Lagrange equations
(see Courant and Hilbert \markcite{CH53} 1953,
p. 296) but can change the boundary terms.  For instance, varying
\begin{equation}
\int_{t_0}^{t_1} \left[ L - \sum_j \frac{d}{dt} \left( q_j
\frac {\partial L}{\partial \dot{q}_j} \right) \right] dt
\label{eq:velaction}
\end{equation}
leads to
\begin{equation}
\int_{t_0}^{t_1} \sum_j \left[ \frac{\partial L}{\partial  q_j} -
\frac{d}{dt} \left( \frac {\partial L}{\partial \dot{q}_j} \right)
\right]  \delta  q_j dt  - \sum_j \left[
q_j \delta \left( \frac{\partial L}{\partial \dot{q}_j} \right)
\right]_{t_0}^{t_1} = 0.
\label{eq:velbdry}
\end{equation}
If the original Lagrangian is quadratic in $\dot{q}_j$,
recovery of the Euler-Lagrange equations requires 
that either $q_j =0$ or $\delta \dot{q}_j =0$ in each boundary term.
In the second case, it is the velocity at the end point which is
the fixed boundary condition, rather than the position.
This raises the possibility
of using a radial velocity, rather than a distance, as the end point
in a cosmological calculation.
In fact Giavalisco et al.  \markcite{G93} (1993) have considered such a
mixed boundary condition, in one place using it to modify a set of 
approximating functions and in another expressing it as a canonical
tranformation of variables.  Their approaches are in practice equivalent
to this one.  Schmoldt \& Saha \markcite{SS98} 
(1998) have succeeded in using a velocity endpoint in their numerical
calculation.

It is straightforward to show
that the boundary term added here does not
change Whittaker's conclusion above.   
Note that the coordinates $q_j$
which provide velocity boundary conditions may be all or only some of
the total number of coordinates $q_i$, as long as the total derivative
is adjusted accordingly. 
   
\subsection{Variable Endpoints}

If the radial velocity is to be used as a form of endpoint, rather
than the (less accurately known) radial distance, the theory of
``free'' endpoints (constrained to move on a manifold of some description)
comes into play.  In addition to the Euler-Lagrange equations, the
solution must now satisfy the
{\em transversality condition} 
which results from the minimization of
the variation at a free end point\footnote{
See Appendix A for detailed formulae.}.
As expressed in Morse's \markcite{Mo34} (1934) notation, this  condition is
\begin{equation}
\left( L - \sum_i \dot{q}_i \frac{\partial L}{\partial \dot{q}_i}
\right) dt^s
+ \sum_i \frac{\partial L}{\partial \dot{q}_i}dq_i^s = 0
\label{eq:transverse}
\end{equation}
where the superscript $s$ denotes a differential taken along
the end manifold ($s$ takes on values designating the initial or
final end points) and $L$ is the integrand.  

Applied to the Lagrangian for a number of bodies moving under their
mutual gravity
\begin{equation}
L = \sum_i m_i \left( \frac{1}{2} \left( \dot{r}^2_i + r_i^2 \dot{\theta}^2_i
+r^2_i \sin^2 \theta_i \dot{\phi}^2_i \right) + G \sum_{j < i}
\frac{m_j}{|{\bf r}_{ij}|} \right)
\label{eq:lagrange}
\end{equation}
and fixing the time, the transversality condition is
\begin{equation}
\sum_i m_i  \left( \dot{r}_i dr^s_i + r_i^2 \dot{\theta}_i d \theta^s_i
+r^2_i \sin^2 \theta_i \dot{\phi}_i d \phi^s_i \right) = 0
\label{eq:xverse1} 
\end{equation}
or more compactly
\begin{equation}
{\bf P} \cdot d{\bf r}^s = 0
\label{eq:xverse2}
\end{equation}
where ${\bf P}$ is the total momentum and $d{\bf r}^s$ any vector
in the end manifold,
a general result constraining the end manifold.  If the velocity action,
expression (\ref{eq:velaction}), is used a modified form of the
transversality condition applies:
\begin{equation}
\sum_i m_i  \left( \dot{r}_i dr^s_i + r_i^2 \dot{\theta}_i d \theta^s_i
+r^2_i \sin^2 \theta_i \dot{\phi}_i d \phi^s_i \right)
- \sum_i m_i \dot{r}_i d r_i^s - \sum_i m_i r_i d \dot{r}_i^s = 0.
\label{eq:velocipede}
\end{equation}
If the end point under consideration has fixed angles and radial velocities,
the left hand side of equation (\ref{eq:velocipede})
vanishes identically.  The velocity-action
transversality condition thus tells us nothing about the manifolds on which
the end-points lie.
At the same time, the velocity action imposes
no additional restrictions on the end manifolds over the position action.

\section{The Second Variation}

We now come to the question of how far the minimization of the action
integral can be used to reproduce the dynamic equations, that is, the
limit of the least-action method strictly defined.  
The limit may be pictured geometrically
by using {\em kinetic foci} as defined by Thompson and Tait \markcite{TT96}
(1896, section
357, p. 428)\footnote{Page and section numbers are identical
in the 1962 Dover reprint.}: ``If, from any one
configuration, two courses differing infinitely little from one
another have again a configuration in common, this second
configuration will be called a kinetic focus relatively to the first:
or (because of the reversibility of the motion) these two
configurations will be called conjugate kinetic foci.''  It can be
shown (for instance, by Whittaker \markcite{W59}
1959, pp. 251-3) that the action is
neither a maximum nor a minimum over a path which includes a pair of
kinetic foci.

More intuitively, if two paths infinitesimally 
close to each other between the same
pair of end points both satisfy the Euler-Lagrange equations, 
the action (along either path) can no longer be a minimum
(and the variation of the variation between them must vanish).

It is easiest to picture kinetic foci using a
toy dynamical problem, that of finding the motion of a ball rolling
on a large sphere, without friction or other complicating effects. 
Geodesics on the spherical surface are paths of least
action in this case.  Clearly there is a unique minimum path for
points close together; that is, if two end points are chosen near each
other, a portion of the great circle joining them gives the least
action.  As the points are taken farther and farther apart
the action 
increases while staying a minimum.  When the points are taken to be opposite
each other, however, there is an infinite number
of solutions all of the same length.  Dynamically, the ball leaves the
starting point and follows a geodesic to the antipode; but if it had left
the starting point at a slightly different angle, it would still pass
through the same antipode.  On a sphere, then,  
kinetic foci are exactly opposite each other.

Still considering the motion of the ball dynamically,
if the trajectory is extended, worse happens.  The path taken, which
passes more than halfway around the sphere, is actually longer than paths
which follow the complement of the great circle.  In fact it is longer
than some small circles.  

Geodesics on a torus provide greater complexity.
Given any two points on the torus there will be a global minimum; a local
minimum path, going the other way around the major radius; and an
infinite series of other local minima, wrapping around one leg or the other
of the torus.  None of these can be continuously varied into another
because of the different number of wrappings.  The action (the length of
the geodesic) tends to infinity as the wrappings increase.

A more mathematically rigorous and useful, but also more complicated,
treatment of the second variation takes us into Morse Theory. 

\subsection{Morse Theory}

Marston Morse \markcite{Mo34}
(1934)  conducted an extensive study of the general
topological properties of variational problems and their solutions.
A summary of some of his results is presented below\footnote{A shorter
and somewhat more accessible presentation of most of Morse's results
is found in Milnor \markcite{M63} (1963).}.

An {\em extremal} is a path which satisfies the Euler-Lagrange
equations.  A {\em critical} extremal is one which makes the action
a minimum.

A function which will play a part in what follows is defined by
\begin{equation}
2 \Omega (s_i, \dot{s}_i) =
\frac{\partial^2 L}{\partial  r_i^2}
s_i^2 + 2
\frac{\partial^2 L}{\partial  r_i \partial \dot{r}_i}
s_i \dot{s}_i +
\frac{\partial^2 L}{\partial \dot{r}_i^2}
\dot{s}_i^2
\end{equation}
for some integrand $L$ and functions $s_i(t)$.
The {\em characteristic form} is\footnote{Here $z$ is used as
a shorthand symbol for the collection of functions $s_i$, and below
it includes also $u_h$ and $u_k$.}
\begin{equation}
Q(z, \lambda) = \int_{t_0}^{t_1}2 \left( \Omega (s_i, \dot{s}_i)
- \lambda s_i s_j \right) dt.
\end{equation}
For a given extremal with fixed end points, the
{\em accessory boundary problem} is defined as
\begin{equation}
\frac{d}{dt} \left( \frac{\partial \Omega}{\partial \dot{s}_i}
\right) - \frac{\partial \Omega}{\partial s_i} + \lambda s_i = 0.
\label{eq:accessory}
\end{equation}
A solution $s_i$ to this equation not identically zero is an
{\em eigensolution} (sometimes {\em eigenvector}) and $\lambda$ an
{\em eigenvalue}.
 The {\em index} of an eigenvalue is the number of linearly independent
eigensolutions corresponding to the eigenvalue.

For free end points the characteristic
form is
\begin{equation}
Q(z, \lambda) = \sum_{h,k} b_{hk} u_h u_k +
\int_{t_0}^{t_1}2 \left( \Omega (s_i, \dot{s}_i)
- \lambda s_i s_j \right) dt
\end{equation}
where $b_{hk}$, $u_h$ and $u_k$ are derived from the second variation
at the end points and are given in Appendix A.  The accessory
boundary problem, equation (\ref{eq:accessory}), stays the same in
form but the solution $s_i$ must now satisfy the transversality condition. 

If the problem is to find the geodesic in a space of a given metric
between two manifolds of some description, the coefficients $b_{hk}$
are a measure of the curvature of the manifolds.
In particular, when the coefficients
vanish the manifolds are flat.

If $\lambda = 0$ the accessory boundary problem becomes
the {\em Jacobi equation} (not to be
confused with the Hamilton-Jacobi equation), which is identical to
the perturbed Euler-Lagrange equation.  If there exist two points
on an extremal at which an eigensolution with eigenvalue
zero vanishes, these points are {\em conjugate points}.
A {\em non-degenerate} extremal is one which has no zero eigenvalues
in the accessory boundary problem.  It is easily shown that conjugate
points (mathematically defined) and kinetic foci (dynamically defined)
are identical.

Determining the least-action limit for a given variational problem
is thus the same as determining the first zero of the Jacobi
function (after the initial point).  This determination is
not generally an easy thing to do.  To take a specific example, 
for a group of bodies moving under
each other's gravity
the Jacobi function ${\bf s}$ satisfies
\begin{equation}
{\bf \ddot{s}}_i = G \sum_{j \not= i} m_j \left(
\frac{3 {\bf r}_{ij}{\bf r}_{ij}}{|{\bf r}_{ij}|^5}
- \frac{1}{|{\bf r}_{ij}|^3} \right) \cdot {\bf s}_i.
\end{equation}
Clearly, solving this is not a convenient way of
finding kinetic foci.  Not only
is this less amenable to integration than the original
dynamic equation, but the original equation must be solved first (which
makes the locating of kinetic foci as a step in solving the dynamical
problem rather pointless).  A more practical method for use in calculation is
called for. 

\subsection{Choquard's Criterion}

Choquard \markcite{C55} 
(1955) studied the motion of bodies in strongly anharmonic potentials
in the context of a semi-classical treatment of Feynman integrals.
He found that multiple solutions to a dynamical problem were possible
through the action of ``forces of reflection'', which allowed indirect
paths from one end point to the other.  In an indirect path, which
corresponds to a stationary rather than a minimum action, at some
time between the end points
\begin{eqnarray}
\frac{d}{dt} \left( T \right)& =& 0\nonumber \\
& =&  \frac{d}{dt} \left(\frac{1}{2m} {\bf p}^2 \right)
\nonumber \\
& = & \frac{1}{m} {\bf p} \cdot \left( - \nabla V \right).
\end{eqnarray}
That is, the momentum must be normal to the force.

To make this reasoning directly applicable to
 the problem at hand, consider a 
solution to the dynamical equations with a given set of end points,
${\bf r}(t)$; it must
conserve total energy $E$, made up of a kinetic part
$T$ and a potential part $V$.  A varied path ${\bf r}(t) + {\bf s}(t)$ 
(where ${\bf s}(t)$ is a Jacobi function) also conserves an energy
$E + \delta E = T + \delta T + V + \delta V$,
and thus the Jacobi function itself conserves $\delta T +\delta V$.  Since
$V$ is a function only of ${\bf r}$, not $\dot{\bf r}$, $\delta V$ is
a function only of ${\bf s}$, and for $|{\bf s}|$ small (which it is by
definition) a linear function.  This means that $\delta V$ reaches
its extreme value when ${\bf s}$ does, and at the same point $\delta T$
has an extremum.  Since $|{\bf s}|$ can vanish only after its maximum,
this point of extremum must occur before a conjugate point.
The extremum of $\delta T$ is given by
\begin{equation}
\frac{d}{dt} \left(\delta T\right) = 0.
\nonumber
\end{equation}
Writing kinetic energy in terms of momentum,
\begin{eqnarray}
T + \delta T & = & \frac{1}{2m} \left( {\bf p} + 
	\delta {\bf p} \right)^2 \nonumber\\
\delta T & \simeq & \frac{1}{m} {\bf p} \cdot \delta {\bf p} \nonumber \\
& = & \frac{1}{m} {\bf p} \cdot \left( - \nabla V \right) \delta t 
\end{eqnarray}
so the condition for an extremum of the variation in kinetic energy is
\begin{eqnarray}
\frac{d}{dt} \left(\delta T\right) & = & 0 \nonumber\\
\frac{1}{m} {\bf p} \cdot \left( - \nabla V \right) & = & 0
\end{eqnarray}
and Choquard's criterion is recovered.  For a situation with
multiple particles, the varied path ${\bf s}$ may be taken to be
different from zero for only one of the particles.  This leads to
the conclusion that
{\em a conjugate point may occur
only after the point where the momentum is normal to the 
force on some body in the system.}  
Keeping in mind the identity of conjugate points and
kinetic foci  as well as Whittaker's result (above),
Choquard's criterion gives a lower bound to the applicability of
the least-action calculation.  Following the trajectory of a dynamic
system from the initial point, it is a minimum of the action at least
until the momentum of some body is normal to the force on that body.
This provides some insight into the shape of
stationary-solution trajectories, as well as (with a further result of
Morse, below) allowing a conclusion to be drawn 
as to the total number of solutions
of all kinds\footnote{Choquard \markcite{C55} 
(1955) notes that his criterion does
not apply to situations in which the trajectory is {\em always} normal
to the acceleration, as in (for example) circular motion.  However,
these situations are generally symmetrical enough to allow the useful
application of Jacobi functions.}.

\subsection{More Morse}

There are several more results from Morse \markcite{Mo34}
(1934) which are of use in the
present problem.  First we require a few more definitions:

A {\em Riemannian} space possesses a positive-definite metric which
can be expressed as a quadratic form:
\begin{equation}
ds^2 = \sum_{i,j} g_{ij} dx^i dx^j.
\label{eq:Riemann}
\end{equation}

The {\em connectivity} $P_k$ of a space\footnote{
This is not to be confused with the {\em connection} of a space,
or whether a space is {\em simply connected} (themselves
distinct topological concepts).} is the
number of distinct homologous families of figures of dimension
$k+1$; that is, within each family one figure can be transformed
into another by a continuous transformation, but a figure in one
family cannot be so transformed into a figure in another.  On a
sphere, for instance, the connectivity $P_0$ is one, since any
line may be transformed into another by a continuous transformation.
On a torus $P_0$ is infinite, since there is an infinite number of
families of curves distinguished from each other by the number of times
they wrap about the large or small radii.

Morse is concerned with the connectivities of the {\em functional
domain}  $\Omega$ of
admissible curves for a given variational problem\footnote{This is
{\em not} the function $\Omega (s, \dot{s})$ found above and in
Appendix A.  The ambiguity in notation is regretted, but it should
not lead to confusion.}, that is those
curves which have the required end points and are continuous along
with their first derivatives.  For the case of a set of trajectories
in three dimensional space it is easiest to consider them
transformed into a
single trajectory in $3n$-dimensional space, between two end points
representing the starting and ending configurations.  Each point in
$\Omega$ represents a trajectory in the $3n$-dimensional space.  
Since no points in $3n$-space are
excluded,
any trajectory can be continuously transformed into any other; so
any point in $\Omega$ can be continuously
transformed into any other.  Any line
in $\Omega$ can then be transformed point by point into any other
line, any plane figure likewise, and so on for all dimensions.
Consequently each connectivity of the space of trajectories
is one.

Morse's important results are:

{\em An extremal which affords a minimum has no negative
 eigenvalues in the associated boundary problem.}
This is equivalent to saying it contains no conjugate points.
Further, {\em the number
of conjugate points of an end point of an extremal $g$ on $g$
is equal to the number of negative eigenvalues in the
associated boundary problem.}

{\em The index of an extremal is the sum of the indices of the
conjugate points of an end point on the extremal.}

{\em The conjugate points of an end point of an extremal $g$
on $g$ form a set of measure zero.}  
This means they are isolated (and thus much easier to
deal with).  More importantly, it means that the probability of
choosing a pair of conjugate points by chance
when setting up the variational problem is essentially zero.

{\em If for a given Riemannian space R and terminal manifold Z
there exists an integral I defined on R such that all
critical extremals are non-degenerate, then the number of
distinct extremals of index $k$ is greater than or equal to
the connectivity $P_k$ of the functional domain $\Omega$.  If the
extremals are of increasing type, the number of extremals
 of index $k$ is equal to the connectivity
$P_k$.}
 
The last is a most useful result.  However, to apply it we must
show that the variational problem meets the requirements.

As demonstrated for example by Whittaker \markcite{W59}
(1959, pp. 247-8, 254)\footnote{See also Arnold (1989), pp. 245ff.} 
the dynamic equations
of a system which has an integral of energy $E$ can be derived
by requiring that the variation of the integral
\begin{equation}
\int 2T dt
\label{eq:euler1}
\end{equation}
(where $T$ is the kinetic energy) vanish, for a fixed value
of $E$.  This formulation is known as Euler's Principle.  
For a system in which the total energy $E$ is the sum of the
kinetic energy $T$ (quadratic in velocities) and potential energy $V$, the
integral (\ref{eq:euler1}) can also be written as
\begin{eqnarray}
I & = & \int 2 \left( E - V \right)^{1/2} \left( T \right)^{1/2} dt
\nonumber \\
 & = & \int 2 \left( E - V \right)^{1/2} \left( a_{ij} \dot{x}^i 
\dot{x}^j \right)^{1/2} dt.
\label{eq:euler2}
\end{eqnarray}
This is
the integral giving arc length on a surface of metric
\begin{equation}
g_{ij} = 4 \left( E - V \right) a_{ij}
\end{equation}
and the space of the possible solutions to our variational problem 
when formulated this way is
indeed Riemannian, with geodesics corresponding to variational solutions.
The cosmological variational problem in proper coordinates 
can be put into this form, so Morse's results apply.
Further, the integral is of increasing type (since kinetic energy
is always positive), so the
number of solutions is given definitely.

Unfortunately, Morse's result  cannot be applied directly 
to Hamilton's principle, as
the functional domain is not Riemannian.  The Lagrangian cannot be put
in the necessary  form of expression (\ref{eq:euler2}), since the potential
energy forms a separate term which is not quadratic in the velocity
differentials.  This means that, if
the preceding theory is to be used, either the problem must be
put in Euler's form or  some
connection of a topological nature must be made between the two
Least Action principles.  The latter is addressed in the next section.

Before leaving Morse Theory, however, it is worthwhile
to note the effect of the end manifolds of the
velocity action on the number
of solutions.  Comparing the characteristic forms and accessory
boundary problems between fixed-point and manifold situations, we find
that the difference lies in the transversality condition and the
quantities $b_{hk} u_h u_k$.  It has already been shown that, using 
Euler's Principle, the transversality condition
holds identically. 

For an initial configuration manifold and Hamilton's Principle, combining two
formulae from Appendix A,
\begin{eqnarray}
b_{hk} u_h u_k& = &  \left[ \left( L - \sum_i \dot{q_i}
\frac{\partial L}{\partial \dot{q_i}} \right) \frac{d^2 t^s}
{de^2} + \left( \frac{\partial L}{\partial t} - \sum_i
\frac{\partial \dot{q_i}}{\partial t} \frac{\partial L}{\partial q_i}
\right) \left( \frac{dt^s}{de} \right)^2 \right. \nonumber \\
&+& \left.
2 \sum_i \left( \frac{\partial L}{\partial q_i} \frac{dt^s}{de}
\frac{dq_i^s}{de} + \frac{\partial L}{\partial \dot{q_i}}
\frac{d^2 q_i^s}{de^2} \right) \right]_1^2 \nonumber \\
& = & \sum_i m_i \left(\dot{r}_i \frac{d^2r^s_i}{d e^2}
 + r_i^2 \dot{\theta}_i \frac{d^2 \theta^s_i}{d e^2}
+r^2_i \sin^2 \theta_i \dot{\phi}_i \frac{d^2 \phi^s_i}{d e^2} \right). 
\end{eqnarray}
Comparing this with equations (\ref{eq:xverse1}) and (\ref{eq:xverse2})
the manifold curvature expression $b_{hk} u_h u_k$ is seen to be 
the dot product of the total momentum vector
with a vector in the end-manifold surface:
\begin{equation}
b_{hk} u_h u_k = {\bf P} \cdot \frac{d^2}{de^2} {\bf x}^s
\end{equation}
which is, near an extremal,
\begin{equation}
b_{hk} u_h u_k = {\bf P} \cdot \Delta {\bf x}^s/e^2.
\end{equation}
For an extremal, the tansversality condition requires that ${\bf P} \cdot
\Delta {\bf x}^s = 0$ (see equation \ref{eq:xverse2}).  More generally,
since ${\bf P}$ is a constant of motion belonging to the solution
(not related to any particular variation around it) and $e$ and
$\Delta {\bf x}^s$ are independently arbitrary, 
${\bf P} \cdot \Delta {\bf x}^s$ cannot vary with $e$, and
thus must vanish (unless $b_{hk} u_h u_k$ is allowed to be
infinite, a pathological case I propose to ignore).
The dot product vanishing causes $b_{hk} u_h u_k$ to vanish
as well.  The same result holds if Euler's Principle is used.

For the velocity action and the final end manifold, with angles and radial
velocities fixed, $b_{hk} u_h u_k$ vanishes identically using either Least
Action Principle.  Thus for each
case {\em the fact that the ends of the action integral lie on manifolds and
not on fixed points is irrelevant to the number of solutions.}  Interpreted
geometrically, the manifolds are flat surfaces.

\section{The Number of Solutions}

\subsection{Catastrophe Theory}

Applying Morse's result on the number of solutions to the problem formulated
using Euler's Principle (which is the only way it is directly
applicable), we find that
there is one minimal extremal, plus one non-minimal
extremal for each integral number of conjugate points.
There are values
of total energy for which there are no solutions.  Most obvious are
 those below the potential
energy of the final end point; those are excluded from the functional
domain at the outset.  For values of the total energy in a gravitational
system which are positive, especially strongly so, there can be only
one solution since the
Jacobi function for a nearly straight-line trajectory never returns to zero; the
saddle-point solutions for these situations can be thought of as occurring
at infinite values of total time.

But the calculation
contemplated (and as performed by Peebles and those following his technique) 
uses Hamilton's Principle.  To
apply Morse to Hamilton a connection must be made between them.  This can
be done by way of the dynamical equations and Catastophe Theory.

Consider a two-dimensional slice of the functional domain $\Omega$, 
the dimensions
being the Eulerian action (time integral of the kinetic energy $T$)
and the Hamiltonian action (time integral of the Lagrangian function $T-V$).  
Choose the slice
so that it contains all the extremals of the problem (see Figure 
\ref{poincare}).   A given value of
total energy will plot as a curve in this slice, with a minimum of 
$\int T dt$ at the
location of the least-action trajectory and other extremals spaced along it
(the latter may show as maxima, minima or points of 
inflection in this plot).  Since the integrand of Euler's Principle is
positive-definite, the least-action solution takes the mimimum time of
all the solutions for a given energy; the saddle-point solutions take increasing
amounts of time for increasing index.  
All solutions will be equilibrium points for the
potential represented by the action.  The slice is thus a
Poincar\'{e} diagram to which Catastrophe
Theory applies, with total energy as the control parameter.   The minimum
solutions correspond to stable equilibria, the saddle-points to unstable
equilibria.

In the same slice plot curves of constant total time.  The index
of a given extremal depends only on the number of 
kinetic foci of the trajectory,
not on the action principle (if any) used to calculate it; in addition, all
extremals are solutions to the dynamic equations.  Thus the minimum of 
$\int L dt$ for each time corresponds to a minimum of $\int T dt$ for
fixed total energy, and the non-minimum extremals will similarly correspond
to non-minimum extremals of the Eulerian action
for other values of total energy.  Again, we have constructed
a Poincar\'{e} diagram (rotated $90^{\rm{o}}$ with respect to the first), 
with total time as the control parameter.

There is at least one least-action solution for a given value of time.  If
there were two (or more), the chain of Eulerian least-action solutions would
have a maximum or a minimum in total time, as shown in Figure \ref{poincare}. 
There are several reasons why this cannot happen; two are outlined below. 

First, viewed as
a Poincar\'{e} diagram in Hamiltonian extremals, Figure \ref{poincare}
requires two chains of
{\em similar} (stable or unstable)
equilibria to meet. 
This sort of topology, a bifurcation without an exchange of stability,
 is forbidden by Catastrophe
Theory; therefore there is only one least-action Hamiltonian extremal.  
Similarly,
there can only be one saddle-point Hamiltonian extremal for each saddle-point
Eulerian extremal\footnote{Expositions of Catastrophe Theory are found in
Lamb \markcite{L32} (1932, sect. 377, pp. 710-12) and Jeans \markcite{J19}
(1919, sect. 18-23, pp. 20-6);
the detailed demonstration of the 
necessity of an exchange of stability is found in in Poincar\'{e} 
\markcite{P85} (1885).}.

Second, note that at a bifurcation point (point C in figure
\ref{poincare})
\begin{equation}
\left| \frac{\partial^2 I}{\partial q_i \partial q_j} \right| = 0
\end{equation}
for the action $I$ and variables $q$ in the two-dimensional slice; 
but this is just the requirement for a degenerate
extremal, to which Morse's results specifically do not apply.  Since, as
noted above, these require some special symmetry in the problem and are
almost impossible to generate by chance, it is reasonable to assume that
our problem does not have them\footnote{It might be possible to exclude
them explicitly from the functional domain $\Omega$, avoiding any
problems at the start.  However, it is conceivable that such an exclusion
would change the topology of $\Omega$ and thus complicate the
question of the connectivities of the space.  For present purposes
it is easier to deny them any place in the problem at the end.}.

We are finally in a position to determine how many solutions there
are to the cosmological variational problem.

If the question is posed in a strictly proper-coordinate,
Newtonian manner it comes out
something like this: given a number of bodies moving under the influence
of each other's gravity, all constrained to occupy the same position at
time zero, and having given positions (or positions in two dimensions,
radial velocities in the third) now; how many possible trajectories are
there?

If the problem is formulated using the Eulerian action (minimum kinetic
energy for fixed total energy), the
space of solutions is Riemannian and the extremals are of increasing type.
There is thus one minimum (and one stationary solution for each number
of kinetic foci).  By way of Catastrophe Theory this is connected to
the Hamiltonian action (the form in which the question is asked above),
which excludes some solutions which require a different value of total time.
{\em There is one minimum solution and a finite number of stationary
solutions.}

For small values of total time the energy will be forced to be 
positive (in order
for the system to get from one configuration to the other, the speeds must be
large, hence the kinetic energy large and positive) and only the least action
solution will appear.  This idea will be expanded below.

Note that if there is no integral of energy these results do not 
apply.  Thus if a calculation attempts to compute the trajectories of a 
number of galaxies in a time-dependent,
external tidal field, or any other case in which
only part of an interacting system is modelled, the number of solutions
cannot be determined from this development\footnote{This does not mean that
Layzer's (1963) cosmic energy equation exempts all interesting distributions
of astronomical objects from the results obtained here.  An integral of
energy still exists for any collection of masses interacting through 
gravity; Layzer's equation only states that a quantity based on comoving
motions and coordinates, which resembles Newtonian energy in some
respects, is not conserved.  Since the number of solutions a problem has
should not depend on which particular variables are used to write it down,
results obtained herein using proper, inertial coordinates apply also to
calculations performed in other ways.}.

\section{A Dynamical Example}

The simplest useful example of a dynamical system in astronomy is the
two-body problem, dealing with a pair of
bodies of reduced mass $M$ 
in an orbit of total energy $E$ and angular momentum $J$.
Imposing a spherical coordinate system ($r, \theta, \phi$)
with the orbit in the plane
of the equator ($\theta = \pi /2$), the trajectory is given
by
\begin{equation}
r = \frac {R_0}{1+e \cos \phi}
\label{eq:basic}
\end{equation}
with $e$ the eccentricity of the orbit and $R_0 = J^2/GM$.
Defining the Jacobi functions in each of the coordinates
as $\delta r = s$, $\delta \theta = \xi$, $\delta \phi = \eta$
and the perturbations in energy and angular momentum as $h$ and $l$
respectively, one eventually finds
\begin{eqnarray}
\xi &=& \xi_0 \sin (\phi - \phi_0) \\
\frac{d \eta}{d \phi} &=& \frac{l}{J} - 2 \frac{s}{r} \\
{\rm for}\; e < 1, \;s &=& \frac{h G M}{2 E^2} \left[ F \sin\phi + e + \left(
	\frac{E l}{J h} \frac{e^2-1}{e} - \frac{e^2+1}{2} \right)
	\cos\phi \right. \nonumber\\
	& & \left. -\frac{1}{2} \frac{e \sin^2 \phi}{1+e \cos \phi} -
	\frac{3 e^2}{\sqrt{1-e^2}} \sin \phi \arctan \left(
	\frac{\sqrt{1-e^2}}{1+e} \tan \frac{\phi}{2} \right) \right] \\
  & &  \nonumber \\
{\rm for}\; e>1, \; s &=& \frac{h G M}{2 E^2} \left[ F \sin\phi + e + \left(
        \frac{E l}{J h} \frac{e^2-1}{e} - \frac{e^2+1}{2} \right)
        \cos\phi \right. \nonumber\\
        & & \left. -\frac{1}{2} \frac{e \sin^2 \phi}{1+e \cos \phi} -
	\frac{3 e^2}{2 \sqrt{e^2-1}} \sin \phi \ln \left|
	\frac{\sqrt{e^2-1} \tan (\phi /2) + 1 + e}
	{\sqrt{e^2-1} \tan (\phi /2) - 1 - e} \right| \right] \\
  & &  \nonumber
\end{eqnarray}
where $F$ is a constant used to adjust the zero point of $s$.
The first expression for $s$ is used for bound (elliptical) orbits,
the second for unbound (hyperbolic).
The practical difficulty of calculations using Jacobi functions
is evident.

The out-of-plane Jacobi function  $\xi$ is, however, simple and gratifyingly
general.  For any eccentricity (indeed, even for unbound trajectories)
conjugate points are found on diametrically 
opposite sides of the orbit.  This
is easy to picture: rotation of the orbit through an infinitesimal
(or even larger) angle
around a line from the orbiting body through the primary, certainly
an allowed variation, leaves the
opposite point unchanged.

For very small $e$, that is for orbits close to circular, $s$ becomes
a simple sine function also, returning to zero after half an
orbit.  For $e \sim 1$, that is for orbits close to a parabola, analysis is
a bit more complicated, though $s$ is approximately sinusoidal and
in no case does $s$ reach zero
again until after half an orbit.  For $e>1$, but not by much, $s$
remains approximately sinusoidal.  For very large $e$ the approximation
is better, that is the  points where $s$ vanishes 
are closer to being $180^{\rm o}$ in longitude
away from each other; but the (unbound)
trajectory might not include enough movement
in longitude to provide a conjugate point for some initial points.

The Jacobi function in longitude, $\eta$, has a behavior which is in 
full even more complicated.  However, note that its derivative is
directly related to $s$.  It can therefore not return to zero until
well after $s$ changes sign.  Among the three Jacobi functions, then,
$\xi$ has its first zero after exactly half an orbit, while the other
two take longer; so
{\em The earliest zero for any perturbation in a two-body system
occurs after half an
orbit, so kinetic foci are $180^{\rm o}$ apart.}

Choquard's criterion is much easier to apply.  There are two points
where the momentum is normal to the gradient of the potential, at
pericenter and apocenter; any pair of conjugate points must lie on
opposite sides of one of these.  Together with Morse's count of
solutions to the Eulerian
variational problem this means that there is an
infinite number of solutions for a given set of end points, one for 
each half-integral number of
revolutions of the orbit.  As noted above, trajectories with energy
lower than the lower of the potential energies of the end points are
excluded from consideration.  Those with positive energy have one
least-action solution and possibly one saddle-point solution at finite
times (depending on whether the first end point is taken far enough
away from perihelion to allow the kinetic focus, $180\arcdeg$ away in
longitude, to appear on the trajectory); the rest at infinite times.

Applied to systems with many bodies, saddle-point
solutions correspond to some sort of multiple-pass trajectory.
If there are two bodies in an orbit that
approximates isolated two-body motion, they can generate kinetic
foci for the whole system.  

Given a bound two-body system with a set of endpoints and a fixed 
time taken to go between them, the minimum-action solution will 
give a trajectory made up of less than half an orbit.  The first 
stationary-action solution will contain more than half an orbit, 
a longer distance, which means a higher speed and thus higher kinetic 
energy.  The second stationary solution will require at least three
times the speed of the minimum solution, thus nine times the kinetic 
energy; few such orbits are bound.  The situation for a many-body
system is rather more complicated, but for most astronomical systems
a significant increase
in the kinetic energy will make total energy positive and
thus the system will become unbound.  In this way the relatively small binding
energy of astronomical systems severely limits the number of saddle-point
solutions (unless there is, say, one or more tightly orbiting pairs of
objects).

\section{Continuum Solutions}

The discrete body approach to galaxy dynamics is of course an
approximation.  It may be justified by the fact that present distances
between galaxies are significantly larger than galaxy dimensions, 
or (more practically) on the basis of our ignorance
of their detailed mass distributions (including such things as dark
matter halos).  But if we are to consider the motions of galaxies
all the way back to their formation it becomes an increasingly bad
approximation, and it would be better to consider a continuous fluid
of gravitating matter.

Indeed, the present picture of galaxy formation has them condensing
out of a smooth fluid.  It would be highly desirable to be able to
follow this process in detail while requiring a certain configuration
as a final end point.  One could investigate, for example, the 
importance of mergers in galaxy dynamics, as well as the problems
encountered by Dunn and Laflamme \markcite{DL95}
(1995) in matching a least-action
calculation to an n-body simulation.

However, in attempting this
we are faced with a massive theoretical complication as
the number of degrees of freedom goes from $3n$ to infinite\footnote{This is
of less practical importance, as a continuum calculation always has
some sort of short-wavelength cutoff (which is addressed in more detail
below).}.  Additional practical difficulty is involved
with the increased complexity of the calculation, using three equations
(continuity, Euler's and Poisson's) instead of one.  However, it can
be done, as Susperreggi and Binney \markcite{SB94}
(1994) have shown (though it tends to be computationally intensive).

Consider, as a first approximation to a continuous-fluid situation, a
large N-body calculation.  Since the results of Morse Theory do not
depend on the number of bodies, there still remains one minimum action
solution and a finite number of stationary action solutions. (The bodies
are now of all the same mass, and are labelled with, say, their ending
coordinates instead of ``M31''; but the Morse-based results are unchanged.)
Adding more bodies increases the resolution of the simulation and the
computational burden, but does nothing to the theory of solutions. Therefore,
so far as a continuous fluid may be considered as made up of discrete masses,
however tiny, there remains one least-action solution and one stationary
solution for each possible value of the index.

\subsection{Orbit-Crossing and Kinetic Foci}

Giavalisco et al. \markcite{G93}
(1993) identified orbit-crossing as a major cause of
multiple solutions, that is, when trajectories from different parts of
the fluid occupy the same point at the same time.  This makes
the mapping of velocity to distance (a major concern of observational
cosmology) multiple-valued.  However, our question---the number of ways
the present velocity and density distribution can arise from the Big
Bang---is different, and orbit-crossing is not necessarily relevant.

To see this, consider a spherically symmetric part of a nearly uniform
universe, of critical density for definiteness.  Suppose that a small
perturbation makes one shell slightly more dense than average and the
shell contained immediately within it less dense.  Over time the dense
shell will expand at a slower rate than the universe as a whole, and
the less dense shell faster; eventually their trajectories will meet, and
there will be orbit-crossing (even with all shells expanding).

To locate the kinetic foci, first write the dynamical equation of a
shell which contains a mass $m(r)$ within a radius $r$:
\begin{equation}
\ddot{r} = - \frac{Gm(r)}{r^2}
\end{equation}
which has the Jacobi equation
\begin{equation}
\ddot{s} = \frac{2 G m(r)}{r^3}s
\end{equation}
which, for shells near critical density, becomes
\begin{equation}
\ddot{s} = \frac{1}{9t^2}s.
\end{equation}
In any spherically symmetric case $s$ can start from zero and go back
to zero only after $r$ changes sign.  In a critical universe (and, indeed,
in any universe before a Big Crunch) this never happens; thus there
are no kinetic foci.  {\em 
Orbit-crossing does not necessarily generate kinetic foci}.

Now consider another nearly uniform universe, but this time allow 
several mass condensations to form.  Place them in such a way as to
generate two binary systems, and allow the tidal torque of each on the
other to send them into bound orbits.  In all this allow none of the
trajectories of mass elements to cross.  After half an orbit kinetic
foci will be generated.  {\em  Kinetic foci do not necessarily generate
orbit-crossing}.

Certainly an orbit-crossing situation in the context of the cosmological
problem demands that two mass elements start in the same place (where
all mass elements start, the origin) and end in the same place (where
their trajectories cross).  At first glance this appears to involve two
trajectories with identical (proper space) endpoints, and thus two solutions
to the equations of motion.  But a solution is made up of all the
trajectories of the bodies included, and whether it is a saddle-point
or a minimum is an attribute of the solution as a whole, not of any
of these bodies.  In fact the two bodies that share end points in an
orbit-crossing situation are two solutions to slightly different
equations of motion, not two different solutions to the same equation.

\subsection{Potential Flow and Kinetic Foci}

A useful simplification, then, for a continuum least-action calculation
would be one that eliminates closed
orbits; that is, one in which there is no
rotation.  Susperreggi and Binney \markcite{SB94}
(1994) used a velocity field derived
from a potential suggested by Herivel \markcite{H55} (1955):
\begin{equation}
{\bf v}(x,y,z,t) = \nabla \alpha (x, y, z, t).
\end{equation}
The field thus derived is both laminar and irrotational; the first term
refers to the fact that it can have no orbit-crossing, and the second to
the fact that it can have no vorticity:
\begin{equation}
\nabla \times {\bf v} = 0
\label{eq:vorticity}
\end{equation}
so they appear to have satisfied all parties.

Unfortunately, it is possible to have rotation in a flow that has no
vorticity.  Equation (\ref{eq:vorticity}) is satisfied by a velocity
field whose longitudinal ($\phi$) component varies inversely with radius, $v_{\phi}
\propto R^{-1}$; Lynden-Bell has pointed this out and, moreover,
shows that it is just the sort of field one expects from tidal 
interactions (Lynden-Bell\markcite{LB96} 1996).  
{\em A velocity field derived from a scalar potential
can generate kinetic foci.}

\subsection{Resolution and Kinetic Foci}

The number of solutions in a continuum calculation thus formally remains
the same, even if the restriction to potential flow is imposed: one
minimum and one or several stationary solutions.  Considering the latter
the situation can appear rather depressing, since any two-body orbit by
any pair of mass-elements, no matter how small, will generate kinetic
foci and thus multiple solutions.  It seems somehow unfair that a
cosmological simulation should lose its minimum status through half the orbit
of its smallest binary star.  In practical terms, this means that a continuum
least-action algorithm which is strictly minimizing will find only one
solution, the one without so much as a half-orbit, which is not necessarily
the right one; while an algorithm which finds all stationary solutions
will find many possible answers, with no clue as to which is more probable.

But cosmological simulations rarely depict single stars.  In practice
there is always a scale below which no detail can be seen; kinetic foci
on this scale cannot affect the minimum status of the calculation.  In
a very simple example, consider a triple star made up of one tight binary
and one wide component.  If all bodies are included, a solution will only
be a minimum through half the orbital period of the close double. However
if the binary is modeled by a single mass, a solution will be a minimum through
half the period of the wide component.  It is a matter of choice which
is the more important trajectory to calculate--or, alternatively, 
whether the
computational burden of calculating several, perhaps many, stationary
solutions is worth maintaining the higher resolution.

In a more complicated situation setting the desirable resolution is 
also more
complicated.  In a rich galaxy cluster, for instance, the dynamical
timescale of the center regions is much shorter than the outskirts, and
varies continuously with radius.
What particular scale is best for the calculation?  The answer is not
obvious.  However, the question is not restricted to least action 
calculations, so it is at least a familiar one.

\section{Summary}

The important results of this study are as follows:

{\em If the action for the cosmological
variational problem can be written in proper
coordinates and an integral of energy exists, there is one minimum
solution.}
Assuming Hamilton's Principle is used,
there may be additional, stationary solutions, one for each number
of kinetic foci, if multiple-pass trajectories exist.  There is a
finite number in total, limited by possible values of energy.
{\em Solutions containing at least one approximately two-body
orbit which passes through more than $180\arcdeg$ in longitude
are not minima.}

{\em Kinetic foci are reached only after the momentum is normal
to the force for some body in the system.}

In so far as a continuous mass distribution may be approximated by an
arbitrarily large number of individual masses,
{\em a continuum least-action calculation also has a single minimum 
solution, but generally a very large number of stationary solutions.} 
These can be limited by setting a lower limit to the resolution of the 
calculation.  The specific size of this resolution may be difficult 
to determine.

{\em A radial velocity, rather than a distance, can be used as an
end point in a numerical variational calculation.}  Forms of the
modified action required have been discovered by Giavalisco et al.
(1993) and used by Schmoldt \& Saha (1998). {\em  Using such an
endpoint has no effect on the number or character of solutions.}

{\em Orbit-crossing is not necessarily related to the number of
solutions of a continuum calculation.} 

\acknowledgements

It is a pleasure to thank Donald Lynden-Bell for drawing my
attention to the least-action problem and suggesting the radial velocity
action.  I received valuable assistance in interpreting topological
ideas from Wendelin Werner and Anthony Quas.  Sverre Aarseth kindly
provided a copy of his n-body code to check 
a sample least-action calculation.  Peter McCoy made many useful
comments on an early version of this paper.
This work has been supported in part by
grants from Herschel Whiting, Marion and Jack Dowell, the British
Schools and Universities Foundation May and Ward Fund, and the University of
Cambridge Institute of Astronomy.

\appendix
\section{Variable End Points}

The following derivations follow Morse \markcite{Mo34}
(1934) with  some
changes in notation and terminology. 
Courant and Hilbert \markcite{CH53} (1953) have a derivation
for the transversality condition which in fact results in the same
formula; however, they require some assumptions about
the end manifold which do not hold in the present situation.

\subsection{The Transversality Condition}

Suppose the problem to be that of minimizing the integral
\begin{equation}
I = \int_{t_1}^{t_2} L \left( q_i,\dot{q_i},t \right) dt
\label{eq:integral1}
\end{equation} 
subject to the condition that one or both of 
the end points are not fixed but must
lie on end manifolds of some description.  
The solution to the problem
is given by  some extremal $g = g(t)$.
Admissible curves for the problem will be those with end points
near those of $g$ and which are continuous along with their first and
second derivatives.  These curves
are described by the $r$ functions $\alpha_h (e)$
such that $\alpha_h (0)$ gives  $g$.  The end points in particular
are given by
\begin{eqnarray}
t^s & = & t^s(\alpha_1, \ldots, \alpha_r) \nonumber \\
q_i^s & =&  q_i^s(\alpha_1, \ldots, \alpha_r) \nonumber \\
s & \in & \left( 1,2 \right) \nonumber
\end{eqnarray}
(the superscript 1 or 2 refers to the initial or final end point).
 Observe
\begin{equation}
q_i^s(\alpha_h (e)) = q_i^s(t^s (\alpha_h (e)), e)
\label{eq:end}
\end{equation}
where $h$ takes on the values 1 to $r$.
 
Integral (\ref{eq:integral1}) is now a function of $e$; considered
this way, the first variation (by Liebnitz' Rule) is
\begin{equation}
I'(e) = \left[ L(t^s) \frac{dt^s}{de} \right]_1^2
+ \int_{t_1(e)}^{t_2(e)} \sum_i \left( \frac{\partial L}{\partial q_i}
\frac{\partial q_i}{\partial e} + \frac{\partial L}{\partial \dot{q_i}}
\frac{\partial \dot{q_i}}{\partial e} \right) dt.
\label{eq:liebnitz}
\end{equation}
After integration by parts and a bit of algebra, one obtains the
Euler-Lagrange equations and
\begin{equation}
\left[ \left( L - \sum_i \dot{q_i}\frac{\partial L}{\partial \dot{q_i}}
\right) \frac{dt^s}{de} + \sum_i \frac{\partial L}{\partial \dot{q_i}}
\frac{d q_i^s}{d e} \right]_1^2 = 0.
\label{eq:transverse1}
\end{equation}
Again, the parametrization by $e$ is arbitrary. 
If $de$ is multiplied out of the above equation,
the normal form
of the {\em transversality condition} is obtained.
If the manifold
on which the end point is allowed to vary is specified
by means of the differentials $dq^s_i$ and $dt^s$, (\ref{eq:transverse1})
contains a condition
fulfilled by the true minimizing end point.  Conversely, 
the transversality condition can sometimes be used
to gain some insight into the end manifold when only the Lagrangian and
the fact of minimization are given.

If the integral to be varied is changed from (\ref{eq:integral1}) to the
velocity action,
\begin{eqnarray}
I^* &=& \int_{t_1}^{t_2} \left( L \left( q_i,\dot{q_i},t \right) 
- \sum_j \frac{d}{dt} \left( q_j \frac{\partial L}{\partial \dot{q_j}} 
\right) \right) dt
\nonumber \\
&=&  I - \left[\sum_j q_j \frac{\partial L}{\partial \dot{q_j}} \right]_1^2
\label{eq:velocity}
\end{eqnarray}
where $j$ denotes those coordinates in which velocity rather than coordinate
is fixed at the end point, the variation of the boundary term must be included
in the transversality condition.  A similar derivation to the above
results in the
{\em velocity-action transversality condition}:
\begin{eqnarray}
\left[ \left( L - \sum_i \dot{q_i}\frac{\partial L}{\partial \dot{q_i}}
- \sum_j q_j \frac{\partial^2 L}{\partial t \partial \dot{q}_j}
\right) \frac{dt^s}{de} \right.  &  + &   \sum_i 
\frac{\partial L}{\partial \dot{q_i}}
\frac{d q_i^s}{de}  
    \nonumber \\
 - \sum_j \left( \frac{\partial L}{\partial \dot{q_j}} + q_j
\frac{\partial^2 L}{\partial q_j \partial \dot{q}_j} \right)
\frac{d q_j^s}{de} & -& \left.  
\sum_j q_j \frac{\partial^2 L}{\partial \dot{q}_j^2}
\frac{d \dot{q}_j}{de} \right]_1^2 = 0.
\label{eq:velocitytransverse}
\end{eqnarray}

\subsection{The Second Variation}

Applying Liebnitz' Rule again gives
\begin{eqnarray}
I''(e)&=& \int_{t_1(e)}^{t_2(e)} \frac{\partial}{\partial e}
\sum_i \left(\frac{\partial L}{\partial q_i} \frac{\partial q_i}
{\partial e} + \frac{\partial L}{\partial \dot{q_i}} \frac{\partial \dot{q_i}}
{\partial e} \right) dt \nonumber \\ &+& 
\left[ \sum_i \left( \frac{\partial L}{\partial q_i} \frac{\partial q_i}
{\partial e} + \frac{\partial L}{\partial \dot{q_i}} \frac{\partial \dot{q_i}}
{\partial e} \right) \frac{dt^s}{de} \right]_1^2 \nonumber \\ &+&
\left[ \frac{\partial}{\partial e} \left( L(t^s) \right) \frac{dt^s}{de}
\right]_1^2 +
\left[ L \frac{d^2 t^s}{d e^2} \right]_1^2.
\label{eq:bigsecond}
\end{eqnarray}
After some algebra this becomes
\begin{eqnarray}
I''(e) & = & \int_{t_1(e)}^{t_2(e)} \sum_i
2 \Omega \left( \frac{\partial q_i}{\partial e},
\frac{\partial \dot{q}_i}{\partial e} \right) dt  +
\int_{t_1(e)}^{t_2(e)} \sum_i \frac{\partial^2 q_i}{\partial e^2}
\left( \frac{\partial L}{\partial q_i}- \frac{d}{dt} \left(
\frac{\partial L}{\partial \dot{q_i}} \right) \right) dt \nonumber \\
&+& \left[ \left( L - \sum_i \dot{q_i}
\frac{\partial L}{\partial \dot{q_i}} \right) \frac{d^2 t^s}
{de^2} + \left( \frac{\partial L}{\partial t} - \sum_i
\frac{\partial \dot{q_i}}{\partial t} \frac{\partial L}{\partial q_i}
\right) \left( \frac{dt^s}{de} \right)^2 \right. \nonumber \\
&+& \left. 
2 \sum_i \left( \frac{\partial L}{\partial q_i} \frac{dt^s}{de} 
\frac{dq_i^s}{de} + \frac{\partial L}{\partial \dot{q_i}}
\frac{d^2 q_i^s}{de^2} \right) \right]_1^2. 
\label{eq:firstsecond}
\end{eqnarray}
The second integral vanishes for extremals.  

While this version
of the second variation is useful,
it may be made a manifestly symmetric quadratic form
in the variations
\begin{equation}
u_h =  \frac{d \alpha_h}{de}. 
\end{equation}
Using these in equation (\ref{eq:firstsecond}) there results  
\begin{eqnarray}
I''(e) & = & \sum_{h,k} \left[ \left( L - \sum_i \frac{\partial q_i}
{\partial t}
\frac{\partial L}{\partial \dot{q_i}} \right) \frac{\partial^2 t^s}
{\partial \alpha_h \partial \alpha_k} + \left( \frac{\partial L}{\partial t} -
\sum_i \dot{q_i} \frac{\partial L}{\partial q_i}
\right) \frac{\partial t^s}{\partial \alpha_h}
\frac{\partial t^s}{\partial \alpha_k} \right.  \nonumber \\
& +& \left. \sum_i \frac{\partial L}{\partial q_i}
\left( \frac{\partial t^s}{\partial \alpha_h}
\frac{\partial q_i^s}{\partial \alpha_k} + 
\frac{\partial t^s}{\partial \alpha_k}
\frac{\partial q_i^s}{\partial \alpha_h} \right) + \sum_i
\frac{\partial L}{\partial \dot{q}_i} \frac{\partial^2 q_i^s}
{\partial \alpha_h \partial \alpha_k} \right]_1^2 u_h u_k \nonumber \\
& + & \sum_h \left[ \left( L - \sum_i \dot{q_i} \frac{\partial L}
{\partial \dot{q_i}}
\right) \frac{\partial t^s}{\partial \alpha_h} + \sum_i 
\frac{\partial L}{\partial \dot{q_i}}
\frac{\partial q_i^s}{\partial \alpha_h} \right]_1^2 
 \frac{\partial^2 \alpha_h}{\partial e^2}  \nonumber \\
&+& \int_{t_1(e)}^{t_2(e)} \sum_i 
2 \Omega \left( \frac{\partial q_i}{\partial e},
\frac{\partial \dot{q}_i}{\partial e} \right) dt  \nonumber \\
&+& \int_{t_1(e)}^{t_2(e)} \sum_i \frac{\partial^2 q_i}{\partial e^2}
\left( \frac{\partial L}{\partial q_i}- \frac{d}{dt} \left(
\frac{\partial L}{\partial \dot{q_i}} \right) \right) dt. 
\label{eq:secondsecond}
\end{eqnarray}

For an extremal satisfying the transversality condition, the
coefficients of $\partial^2 \alpha_h / \partial e^2$ as well as the
last integral vanish; and
we are left with the second variation integral, as in the case of
fixed end points, and a symmetrical quadratic form in the variations
at the end points.  The variations at the end points and within
the integral are related by
\begin{eqnarray}
\frac{\partial q_i}{\partial e}& = & \sum_h \left[ 
 \frac{\partial q_i}{\partial \alpha_h}  -
\dot{q}_i  \frac{\partial t}{\partial \alpha_h}
 \right] u_h \nonumber \\
\frac{\partial \dot{q_i}}{\partial e}& = & \sum_h \left[
\frac{\partial \dot{q_i}}{\partial \alpha_h} -
\ddot{q}_i  \frac{\partial t}{\partial \alpha_h}
\right] u_h
\end{eqnarray}
where evaluation is carried out at the end points.  Morse defines the
quantities $b_{hk}$ for an extremal satisfying the transversality
condition via
\begin{equation}
I''(0) = \int_{t_1(e)}^{t_2(e)} \sum_i
2 \Omega \left( \frac{\partial q_i}{\partial e},
\frac{\partial \dot{q}_i}{\partial e} \right) dt
+\sum_{h,k} b_{hk} u_h u_k
\end{equation}
and uses this notation for his definitions of the index form.

If the velocity action is used, the second variation of the boundary
term must be calculated and added to the expression above. 
Following the lines of the above derivation
one finds
\begin{eqnarray}
I^*{}''(e) & = &  \int_{t_1(e)}^{t_2(e)} \sum_i
2 \Omega \left( \frac{\partial q_i}{\partial e},
\frac{\partial \dot{q}_i}{\partial e} \right) dt  +
\int_{t_1(e)}^{t_2(e)} \sum_i \frac{\partial^2 q_i}{\partial e^2}
\left( \frac{\partial L}{\partial q_i}- \frac{d}{dt} \left(
\frac{\partial L}{\partial \dot{q_i}} \right) \right) dt \nonumber \\
&+& \left[ \left( L - \sum_i \dot{q_i}
\frac{\partial L}{\partial \dot{q_i}} 
- \sum_j q_j \frac{\partial^2 L}{\partial t \partial \dot{q}_j}
\right) \frac{d^2 t^s}
{de^2} + \left( \frac{\partial L}{\partial t} - \sum_i
\frac{\partial \dot{q_i}}{\partial t} \frac{\partial L}{\partial q_i}
- \sum_j q_j \frac{\partial^3 L}{\partial t^2 \partial \dot{q_j}}
\right) \left( \frac{dt^s}{de} \right)^2 \right. \nonumber \\
&+& 
2 \sum_i \left( \frac{\partial L}{\partial q_i} \frac{dt^s}{de}
\frac{dq_i^s}{de} + \frac{\partial L}{\partial \dot{q_i}}
\frac{d^2 q_i^s}{de^2} \right) -\sum_j 2 \left( 
\frac{\partial^2 L}{\partial t \partial \dot{q_j}} + q_j
\frac{\partial^3 L}{\partial t \partial q_j \partial \dot{q_j}}
\right) \frac{d q_j}{d e} \frac{d t^s}{d e} - \nonumber \\
&-& \sum_j \left( \frac{\partial L}{\partial \dot{q_j}}
+ q_j \frac{\partial^2 L}{\partial q_j \partial \dot{q_j}}
\right) \frac{d^2 q_j^s}{d e^2} - \sum_j q_j \frac{\partial^2 L}
{\partial \dot{q_j}^2} \frac{d^2 \dot{q}_j^s}{d e^2}
- \sum_j q_j \frac{\partial^3 L}{\partial \dot{q_j}^3}
\left( \frac{ d \dot{q}_j^s}{d e} \right)^2 \nonumber \\
&-& \sum_j \left( 2 \frac{\partial^2 L}{\partial q_j \partial \dot{q_j}}
+ q_j \frac{\partial^3 L}{\partial q_j^2 \partial \dot{q_j}} \right)
\left( \frac{d q_j^s}{d e} \right)^2 - \sum_j \left( 2 
\frac{\partial^2 L}{\partial \dot{q_j}^2}
+ q_j \frac{\partial^3 L}{\partial q_j \partial \dot{q_j}^2} \right)
\frac{d q_j^s}{d e} \frac{d \dot{q}_j^s}{d e} \nonumber \\
&-& \left. \sum_j 2 \frac{\partial^3 L}{\partial t \partial \dot{q_j}^2}
\frac{d \dot{q}_j^s}{d e} \frac{dt^s}{de}
\right]_1^2.
\end{eqnarray}

Again, a symmetric form may be found for extremals which satisfy the
transversality condition.  Following the above derivation, one obtains
\begin{eqnarray}
I^*{}''(0) & = & \sum_{h,k} \left[ \left( L - \sum_i \frac{\partial q_i}
{\partial t}
\frac{\partial L}{\partial \dot{q_i}} 
- \sum_j q_j \frac{\partial^2 L}{\partial t \partial \dot{q_j}}
\right) \frac{\partial^2 t^s}
{\partial \alpha_h \partial \alpha_k} \right. \nonumber \\
&+& \left( \frac{\partial L}{\partial t} -
\sum_i \dot{q_i} \frac{\partial L}{\partial q_i}
- \sum_j q_j \frac{\partial^3 L}{\partial t^2 \partial \dot{q_j}}
\right) \frac{\partial t^s}{\partial \alpha_h}
\frac{\partial t^s}{\partial \alpha_k}   \nonumber \\
& +& \sum_i \frac{\partial L}{\partial q_i}
\left( \frac{\partial t^s}{\partial \alpha_h}
\frac{\partial q_i^s}{\partial \alpha_k} +
\frac{\partial t^s}{\partial \alpha_k}
\frac{\partial q_i^s}{\partial \alpha_h} \right) + \sum_i
\frac{\partial L}{\partial \dot{q}_i} \frac{\partial^2 q_i^s}
{\partial \alpha_h \partial \alpha_k}  \nonumber \\
&-& \sum_j \left( \frac{\partial L}{\partial \dot{q}_j}
+ q_j \frac{\partial^2 L}{\partial q_j \partial \dot{q_j}} \right)
\frac{\partial^2 q_j^s}
{\partial \alpha_h \partial \alpha_k}
-\sum_j q_j \frac{\partial^2 L}{\partial \dot{q_j}^2}
\frac{\partial^2 \dot{q}_j^s}
{\partial \alpha_h \partial \alpha_k} \nonumber \\
&-& \sum_j 2 \left( \frac{\partial^2 L}{\partial t \partial \dot{q_j}}
+ q_j \frac{\partial^3 L}{\partial t \partial q_j \partial \dot{q_j}}
\right) \frac{\partial q_j^s}{\partial \alpha_h}
\frac{\partial t^s}{\partial \alpha_k} - \sum_j q_j
\frac{\partial^3 L}{\partial \dot{q}_j^3} \frac{\partial q_j^s}{\partial \alpha_h}
\frac{\partial \dot{q}_j^s}{\partial \alpha_k} \nonumber \\
&-& \sum_j \left( \frac{\partial^2 L}{\partial q_j \partial \dot{q_j}}
+ q_j \frac{\partial^3 L}{\partial q_j^2 \partial \dot{q_j}}
\right) \frac{\partial q_j^s}{\partial \alpha_h}
\frac{\partial q_j^s}{\partial \alpha_k}
- \sum_j \left( \frac{\partial^2 L}{\partial \dot{q_j}^2}
+ q_j \frac{\partial^3 L}{\partial q_j \partial \dot{q_j}^2}
\right) \frac{\partial q_j^s}{\partial \alpha_h}
\frac{\partial \dot{q}_j^s}{\partial \alpha_k} \nonumber \\
&-& \left. \sum_j 2 q_j \frac{\partial^3 L}{\partial t \partial \dot{q_j}^2}
\frac{\partial \dot{q}_j^s}{\partial \alpha_h}
\frac{\partial t^s}{\partial \alpha_k}
\right]_1^2 u_h u_k \nonumber \\
&+&\int_{t_1}^{t_2} \sum_i
2 \Omega \left( \frac{\partial q_i}{\partial e},
\frac{\partial \dot{q}_i}{\partial e} \right) dt \\
&=& \sum_{h,k} b_{hk}u_h u_k + \int_{t_1}^{t_2} \sum_i
2 \Omega \left( \frac{\partial q_i}{\partial e},
\frac{\partial \dot{q}_i}{\partial e} \right) dt.
\end{eqnarray}

\clearpage

\figcaption[poincare.eps]{Poincar\'{e} diagram connecting the two
Least Action Principles.  For a given set of end points (or manifolds)
a slice of the functional domain, which by construction contains all
the solutions to the problem, is plotted.  The Hamiltonian action
is the vertical
coordinate, the Eulerian action the horizontal coordinate.
For a given value of total
energy (the Eulerian control parameter; say, $E_1$, $E_2$ or $E_3$)
there will be a single minimum
solution and a series of stationary solutions.  For a given value of
total time (the Hamiltonian control parameter, say $t_1$, $t_2$ or $t_3$)
there will be at least
one minimum solution.  In terms of Catastrophe Theory, the minimum
solutions form a chain of stable equilibria (shown as a solid line in
the figure), the stationary solutions
a chain of unstable equilibria (not shown in this figure for clarity).
If there were two Hamiltonian solutions
on any Eulerian branch of solutions, as shown here, it would require
the meeting of two chains of Hamiltonian similar equilibria (at point
C) without
an exchange of stability.  Such a situation is forbidden by Catastrophe
Theory, as described in the text.  \label{poincare}}


\begin{references}

\reference{A89} Arnold, V. I. 1989, Mathematical Methods of Classical Mechanics, 2nd. ed. (New York: Springer-Verlag, Inc.) \\
\reference{B60} Bondi, H. 1960, Cosmology, 2nd. ed. (Cambridge: Cambridge University Press); see especially chapter IX \\
\reference{C55} Choquard, P. 1955, Helvetica Physca Acta, 28, 89 \\
\reference{CH53} Courant, R., \& Hilbert, D. 1953, Methods of Mathematical Physics, Volume I (London: Wiley-Interscience) \\
\reference{DL95} Dunn, A. M., \& Laflamme, R. 1995, \apjl, 443, L1 \\
\reference{H55} Herivel, J. W. 1955, Proceedings of the Cambridge Philosophical Society, 51, 344 \\
\reference{G93} Giavalisco, M., Mancinelli, B., Mancinelli, B. J. \& Yahil, A. 1993, \apj, 411, 9 \\
\reference{J19} Jeans, J. H. 1919, Problems of Cosmogony and Stellar Dynamics (Cambridge: Cambridge University Press) \\
\reference{L32} Lamb, H. 1932, Hydrodynamics, sixth edition (Cambridge: Cambridge University Press) \\
\reference{L63} Layzer, D. 1963, \apj, 138, 174 \\
\reference{LB96} Lynden-Bell, D. 1996, Current Science, 70, 789 \\
\reference{M63} Milnor, J. 1963, Morse Theory  (Princeton: Princeton University Press) \\
\reference{Mo34} Morse, Marston 1934, The Calculus of Variations in the Large (New York: American Mathematical Society) \\
\reference{P80} Peebles, P. J. E. 1980, The Large-Scale Structure of the Universe (Princeton: Princeton University Press) \\
\reference{P89} Peebles, P. J. E. 1989, \apjl, 344, L53 \\
\reference{P90} Peebles, P. J. E. 1990, \apj, 362, 1 \\
\reference{P94} Peebles, P. J. E. 1994, \apj, 429, 43 \\
\reference{P85} Poincar\'{e}, H. 1885, Acta Mathematica, VII, 259 \\
\reference{SS98} Schmoldt, I. \& Saha, P. 1998, \aj, 115, 2231 \\
\reference{SB94} Susperregi, M., \& Binney, J. 1994, \mnras, 271, 719 \\
\reference{TT96} Thompson, Sir William (Lord Kelvin), \&  Tait, P. G. 1896, A Treatise on Natural Philosophy, Volume 1 (Cambridge: Cambridge University Press); revised and published in 1912 as Principles of Mechanics and Dynamics, also by Cambridge University Press; the latter reprinted 1962 (New York: Dover).  \\
\reference{V93} Valtonen, M. J., Byrd, G. G., McCall, M. L., \& Innanen, K. A. 1993, \aj, 105, 886 \\
\reference{W59} Whittaker, E. T. 1959, A Treatise on the Analytical Dynamics of Particles and Rigid Bodies, 4th edition (Cambridge: Cambridge University Press) \\
\end{references}
\end{document}